# A Deep Learning-Based Workflow for Detection of Lung Nodules in Chest Radiograph

YANG TAI, YU-WENG FANG*, FANG-YI SU AND JUNG-HSIEN CHIANG

*Department of Computer Science and Information Engineering*
*National Cheng Kung University*
*Tainan, 701 Taiwan(ROC)*

*equal contribution

**PURPOSE**: This study aimed to develop a deep learning-based tool to detect and localize lung nodules with chest radiographs(CXRs). We expected it to enhance the efficiency of interpreting CXRs and reduce the possibilities of delayed diagnosis of lung cancer.

**MATERIALS AND METHODS**: We collected CXRs from NCKUH database and VBD, an open-source medical image dataset, as our training and validation data. A number of CXRs from the Ministry of Health and Welfare(MOHW) database served as our test data. We built a segmentation model to identify lung areas from CXRs, and sliced them into 16 patches. Physicians labeled the CXRs by clicking the patches. These labeled patches were then used to train and fine-tune a deep neural network(DNN) model, classifying the patches as positive or negative. Finally, we test the DNN model with the lung patches of CXRs from MOHW.

**RESULTS**: Our segmentation model identified the lung regions well from the whole CXR. The Intersection over Union(IoU) between the ground truth and the segmentation result was 0.9228. In addition, our DNN model achieved a sensitivity of 0.81, specificity of 0.82, and AUROC of 0.869 in 98 of 125 cases. For the other 27 difficult cases, the sensitivity was 0.54, specificity 0.494, and AUROC 0.682. Overall, we obtained a sensitivity of 0.78, specificity of 0.79, and AUROC 0.837.

**CONCLUSIONS**: Our two-step workflow is comparable to state-of-the-art algorithms in the sensitivity and specificity of localizing lung nodules from CXRs. Notably, our workflow provides an efficient way for specialists to label the data, which is valuable for relevant researches because of the relative rarity of labeled medical image data.

# Introduction

Lung cancer is one of the leading causes of cancer deaths worldwide. According to the WHO, there were 2.21 million new cases of lung cancer and 1.8 million lung cancer deaths in 2020, making it the most deadly cancer worldwide. One of the main reasons that make it so deadly is that it is often insidious. In fact, a large proportion of lung cancer patients are diagnosed incidentally and in more progressive stages. Treating lung cancer in such conditions is usually challenging because of metastases, poorer physical condition of patients, and other issues. Although new treatments[1] such as target therapy, immunotherapy, and chemotherapy are available, the outcomes of progressive lung cancer still leave something to be desired. The outcomes of cancers are generally better if diagnosed and treated at an earlier stage. Therefore, early detection is undoubtedly a key to improving the outcome of lung cancer.

One of the easiest ways to detect cancer early is regular screening. Regular screening is part of the health insurance policy for certain types of cancer in Taiwan[2]. Currently, low-dose computed tomography is reliable in screening lung cancer. However, low-dose CT screening for lung cancer is not subsidized in Taiwan, partly because of the health risks related to its higher radiation exposure. A favorable solution to screening lung cancer is CXR. CXR is one of the most frequently prescribed medical imaging examinations because of its convenience and clarity. One of the biggest problems of using CXR to screen lung cancer is its sensitivity in detecting lung nodules or tumors.

Deep learning algorithms such as DNNs have proven their abilities in identifying specific lesions in X-ray images. Specifically, some researchers have proposed robust models in identifying lung tumors or nodules from CXRs. Some focus on classifying the CXRs as positive or negative of lung nodules, such as Darknet-53[4], and some even managed to localize the nodules. In recent studies, segmentation models for localizing lung nodules with CXRs have outstanding performances[5][6]. In these studies, careful labeling of the data is usually necessary, which could be a laborious task for the labelers. Thus, in addition to a deep learning-based workflow for localizing lung nodules, we also proposed an efficient and reliable labeling system.

# Materials And Methods

## Training Dataset Composition

Our train dataset is composed of two independent datasets. One of them comes from the NCKUH database. In total, 238 cases with a confirmed lung cancer diagnosis were included as our training data. In addition, 619 cases from the publicly-available sources, the VBD[7] dataset, were also included as part of our training dataset. Cases with combined lung disease were excluded. Finally, we included 857 cases as training data. All CXRs were obtained posteroanterior by using a digital radiography system. Two surgeons of the Division of Chest Surgery at NCKUH were responsible for labeling the data. The annotation result was confirmed by reviewing the chest CT scan and its report of the same patient. We also excluded inadequate cases with the following conditions:

1. The interval between the chest CT scan and CXR is longer than two weeks.
2. The CXR is taken by a portable radiography system.
3. Poor image quality or inadequate positioning

## Patch-Based Training

We presented a two-staged workflow here. First, we identified the pulmonary region of each CXR with a segmentation model. We applied transfer learning by implanting pretrained weights of ImageNet[8] into our model. After identifying the lung region in a CXR, we split it into 16 patches, with eight patches at each side. We normalized the contrast of the patches with ImageNet normalization parameters[9]. Afterward, we augmented our datasets by applying transformations such as adjustments of contrast and brightness, horizontal flips, and rotations[10]. Subsequently, we adopted a ResNet[11]-34 model with pretrained weights of ImageNet and trained the model with our labeled patches. All patches were resized to 224 by 224 with the width-height ratio preserved. If the patch is not square originally, we pad black(rgb(255, 255, 255)) at the shorter side. For our task's purpose, we switched the last linear layer with two out_features, positive and negative.

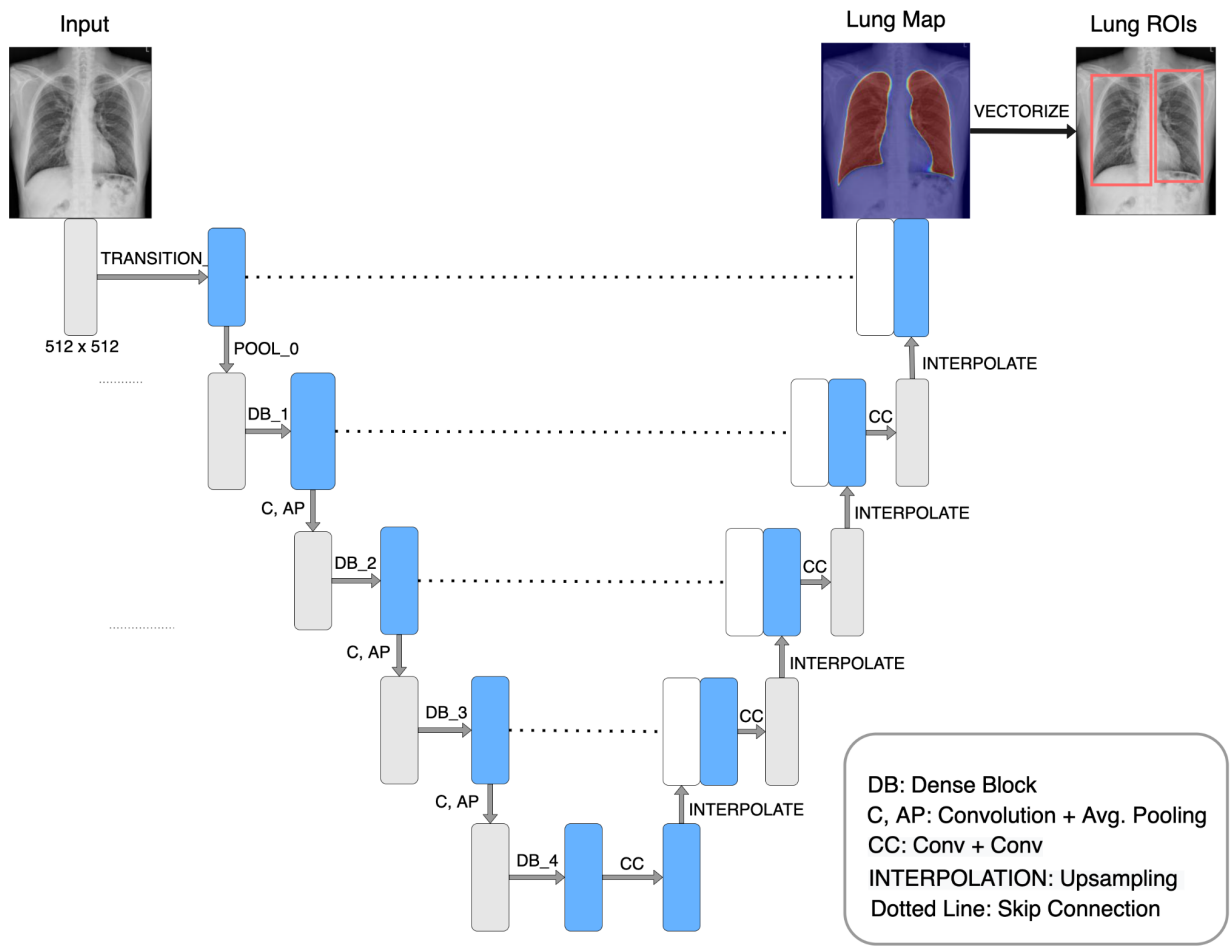

Figure 1: Overview of the lung segmentation model

## Segmentation Stage

We built our segmentation model[12] for identifying the lung regions of CXRs in our all datasets as lung regions of interest(ROIs), as shown in Figure 1. We constructed our model with a U-Net[13]-based architecture and DenseNet-161[14] network as the encoder. Our model first outputs a mask with the shape of the lung. Consequently, we transformed the output to bounding boxes and then sliced the patches out of CXRs according to the bounding boxes.

We sliced the lung regions into 16(4 by 4) patches. Each side of the lung includes eight patches, as shown in Figure 2b. The patches are overlapped with their neighbors(horizontal and vertical), as shown in Figure 2a. This overlapping pattern prevents the model from missing the nodules near the border of the patch or, conversely, classifying two patches as positive by the identical nodule.

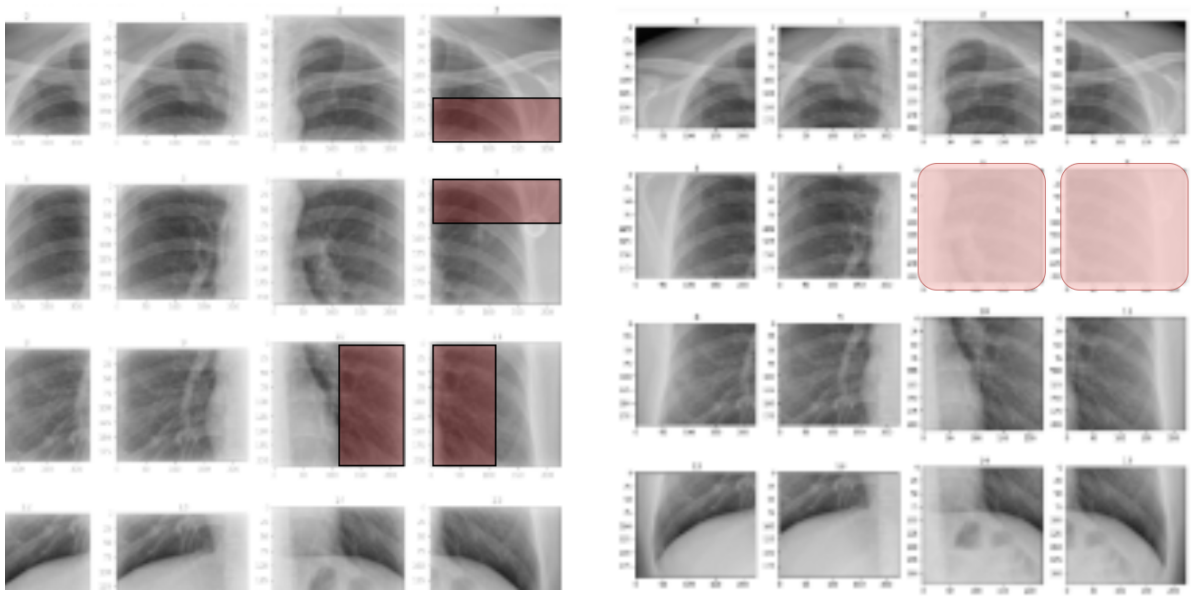

Figure 2 a / b:
a. The pattern of overlap between patches; the overlapped areas are colored by red.
b. The presentation of the patches; the labeled areas are colored by pink.

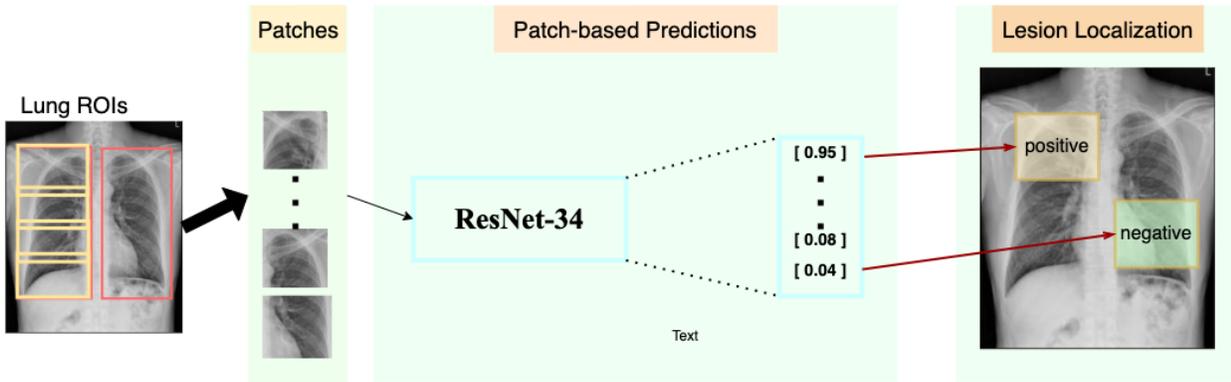

Figure 3: Cropping patches from the lung region and classifying the patches

## Labeling

In an earlier stage of our study, we did not slice the CXRs into patches. We labeled the nodules readily from the whole CXR. Thus, when we adopted the patch-based method later, we transformed the form of the labeling. Specifically, we labeled the patches as positive if they intersected with the labeled nodules. If a nodule lies on more than one patch, we give the positive label to the patch that has the largest intersecting area with the labeled nodule. (Note: a nodule could lie on more than two patches.) Later in our research, we adopted a more efficient way to label the CXRs. We labeled the CXRs simply by clicking on the patches with nodules. The process is pretty like the image recognition reCAPTCHA test. (Figure 2b)

## Classification Stage

In the second stage, the processed image patches are input to a classifier and classified as positive or negative. We used the residual neural network of ResNet-34 to complete the task(Figure 3).

First, we split our dataset into the training set and the validation set by a 3 to 1 ratio randomly[15]. The training set has 12688 patches, and the validation set has 4144 patches. We adopted a mini-batch[16] gradient descent, with a mini-batch size of 32. The gradient descent algorithm minimizes the Weighted Cross Entropy Loss[17]. We set the learning rate by 0.001 for the first 20 warm-up epochs and apply a cosine annealing function to the learning rate[18]. The model generates the possibility of each patch that there is a nodule. We set the positive threshold as 0.9; if the positive possibility is greater than 0.9, it is classified as positive, else negative.

## Test

Our test dataset comprised 141 CXRs from the MOHW database between 2018 and 2020. All of these cases have a confirmed diagnosis of lung cancer. We excluded 16 images for nonstandard view or poor interpretability and labeled the remaining 125 CXRs. The images were also cut into 16 patches and labeled. We inspected the CT scan and its report to ensure the correctness of the labeling. Notably, we marked 27 out of 125 CXRs as difficult cases.

# Results

## Segmentation

We tried to build our segmentation model with UNet and UNet++[19] as the basis of architecture. Our segmentation model outlined bilateral lungs with an ideal IoU of 0.9228(Table 1) in the end. This result is precise enough for slicing the patches without losing the lung region in the CXRs.

| Model | Backbone(Encoder) | IoU |
|-------|-------------------|-----|
| U-Net++ | ResNet 34 | 0.9181 |
| U-Net++ | ResNet 50 | 0.9121 |
| U-Net | DenseNet-161 | **0.9228** |

Table 1: Segmentation model: The performances of a few combinations.

In the second stage, our model classified the patches as positive or negative. We used ROC curve to evaluate the performance of the classifying model. However, smaller patch sizes caused class imbalance in our task, with negative patches much more than positive. Thus, we also calculated the AUPR of the model since AUPR expresses the susceptibility of classifiers to imbalanced datasets and allows for an accurate and intuitive interpretation of their performances[20].

| Model | Dataset | AUROC-val | AUPR-val | AUROC-train | AUPR-train |
|-------|---------|-----------|----------|-------------|------------|
| ResNet-34 | NCKUH + VBD | 0.850 | 0.6689 | 0.922 | 0.838 |
| ResNet-34 | NCKUH | 0.786893 | 0.54282 | - | - |

Table 2: Classification model: The performances of using different sets of data in training

Among all 12688 patches(793 CXRs), 3808 of them come from the NCKUH dataset (238 CXRs) and 8880 from the VBD dataset(555 CXRs). The validation set consists of 4144 patches, with 1152 from the NCKUH dataset(72 CXRs) and 2992 from the VBD dataset(187 CXRs). We trained the model in two settings: the NCKUH and VBD datasets together or NCKUH datasets alone. The mix of NCKUH and VBD datasets yielded a better result(Table 2). The area under receiver operating characteristics(AUROC) is 0.850, and area under precision-recall curve (AUPR) is 0.6689.

We also tested a 6-patch approach, splitting the lung regions into 6 patches, three in a column for each lung.

The 6-patch approach has an AUPR of 0.720, slightly better than the 16-patch approach. Meanwhile, it has an AUROC of 0.820, slightly worse than the 16-patch approach(Table 3). In the end, we adopted the 16-patch approach.

| Model | Dataset | # of Patch | AUROC-val | AUPR-val |
|-------|---------|-----------|-----------|----------|
| ResNet-34 | NCKUH + VBD | 16 | 0.850 | 0.6689 |
| ResNet-34 | NCKUH + VBD | 6 | 0.820 | 0.720 |

Table 3: The performance of the 6-patch and 16-patch approach

## Visualization

Additionally, we apply Class Activation Map(CAM)[21] to demonstrate the activation maps of our model, see if the model focuses on the right spots when it makes the decision. Focusing on the right locations is as important as the correctness metrics for a classification model. As shown in Figure 4, our model focuses on the correct area where the nodules are located.

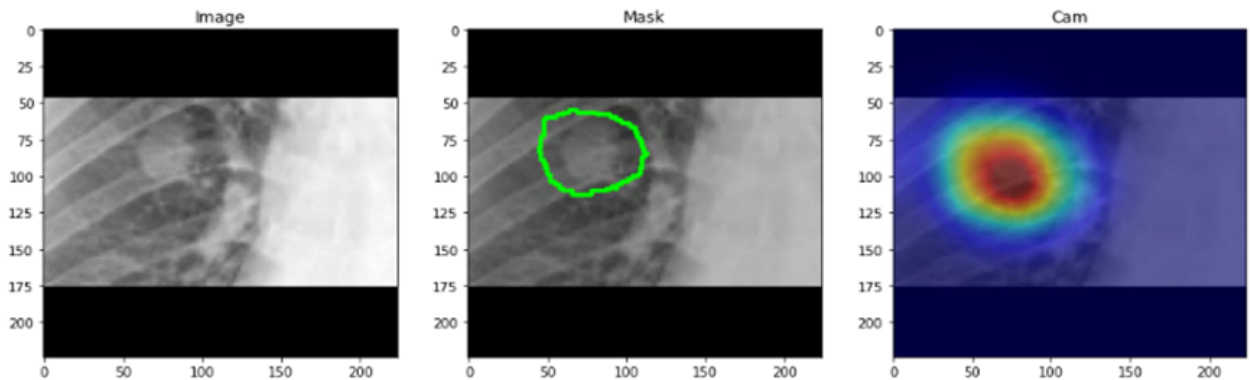

Figure 4: Visualization of the model's attention.

## Test

The test results are shown in Table 4. Our two-staged model achieved the AUROC of 0.837, with a sensitivity of 0.780 and specificity of 0.790. The performance on the same dataset without difficult cases is even better. The AUROC in that condition is 0.869, with a sensitivity of 0.810 and specificity of 0.822. Specifically for those difficult cases, the AUROC is 0.682, with a sensitivity of 0.54 and specificity of 0.494.

|  | # of Case | AUROC | Sensitivity | Specificity |
|---|---|---|---|---|
| All Cases | 125 | 0.83705 | 0.780 | 0.790 |
| w/o Difficult Cases | 98 | 0.86927 | 0.810 | 0.822 |
| Difficult Cases only | 27 | 0.68199 | 0.54 | 0.494 |

Table 4: Performance on the test set. The definition of the difficult case will be present in Discussion.

# Discussion

We developed a workflow composed of two DNNs. It provides a reliable and efficient way to present the nodules in CXRs for clinical application.

Initially, we split the lung region on a CXR to 6 patches. It turns out that the 6-patch approach is slightly better than the 16-patch approach in AUPR. However, the 16-patch approach performed better than the 6-patch approach in the AUROC. Moreover, 16-patch of lung region is also more practical for clinical application. Thus, we adopted the 16-patch approach instead of the 6-patch dataset.

It is challenging for human specialists to detect a nodule at specific locations such as the apices of lungs, the intersecting shadows of ribs, and the retrocardiac region. In such cases, we annotated the CXR as a difficult case. In general cases, it takes our surgeons 1 to 3 minutes to complete the labeling of a CXR. However, in difficult cases, it generally takes them 10 to 20 minutes to complete labeling on that CXR.

We would like to know if there is a nodule in the lungs and if positive, we want to locate the nodule instantly. Our workflow can achieve the goals. Furthermore, labeling medical data is costly because only a few specialists, such as radiologists, can label them, and our workflow provides an efficient way of labeling the data. Therefore, our workflow is also promising for other medical applications, such as detecting other types of lesions in CXRs and detecting lesions or abnormalities in other types of standardized medical images such as KUB.

According to the features mentioned above, our model is helpful in medical facilities. Specifically, it is instrumental in the following circumstances:

1. **Massive amount of incoming data**

When we obtain CXRs in a large amount with only a few specialists available, the order of inspecting them matters. It is quite common that images are inspected a few weeks after being taken. Such a condition may cause delayed diagnosis of lung cancer. However, our model can sort the data by the risk lung nodules are present. In other words, it provides the specialists an order to deal with the massive medical data. As a result, our workflow may reduce the possibility of delay in diagnosing lung cancer.

2. **Mobile medical system:**

Our model also makes mobile medical services more beneficial. Since mobile medical services usually include CXR, it is highly available to people in remote areas. When medical service vehicles or stations collect them from people living in such areas, the requirement of efficiently processing these data becomes significant. Our system can process these data, deciding the priorities for specialists to inspect them.

3. **Screening of lung cancer:**

Last but not least, as we mentioned in the beginning, though low-dose CT is sensitive enough for screening lung cancer, it has significant shortcomings such as radiation exposure. On the other hand, CXR is not used for screening lung cancer because of lower sensitivity and specificity. Since our model improved both of them, it is possible to use CXR as a screening tool for lung cancer in the future.

## Limitations

There are several limitations to our study. First, our data came with a restricted diversity. We do not have sufficient information to evaluate the disparities between different racial or geological groups. Thus, the generality of our model might be of concern. Second, we did not take the relationship between patches into account to optimize the performance of our classifier. Since we know that 16 patches represent a single CXR, we might improve the performance of our model with this information. Third, we sliced the patches with the resized CXRs, which are of lower resolution compared with the original CXRs. Ideally, we should map the segmentation result back to the original images and then slice patches with the original images. It would produce patches with better resolution and might bring about better performance. Fourth, we did not measure and record how much time the surgeons spent labeling each CXR. The information helps estimate the amount of time saved by using our workflow. Fifth, we knew those were all positive cases when we labeled the CXRs from NCKUH and MOHW datasets. However, our model did not have such information at every stage, which made comparing human and AI model performance impossible in our study.

Lastly, the false positive(FP) and false negative(FN) rates are 21% and 22%, respectively. Though comparable to prior studies, they should be improved to be more reliable. Such applications in medicine are generally demanding in performance. We believe a more extensive and diverse dataset and further studies in adjusting the workflow can further improve the performance.

# Acknowledgments

*We acknowledge the dedication of Dr. Yau-Lin Tseng, Dr. Wei-Li Huang, and Dr. Chen-Yu Wu of the Chest Surgery Division of the Surgery Department, National Cheng Kung University Hospital, in this research. In addition, we also acknowledge the Ministry of Health and Welfare for providing data from the national database used in this study.*